\documentclass[prl,aps,tightenlines]{revtex4}

\usepackage[]{graphicx}

\usepackage{amsfonts}%
\usepackage{amsmath}%
\usepackage{bm}
\usepackage{graphicx}
\def\be#1{\begin{equation}\label{#1}}
\def\ee{\end{equation}}
\def\bea#1{\begin{eqnarray}\label{#1}}
\def\eea{\end{eqnarray}}

\def\expm2piOmega{e^{-2\pi\Omega}}

\begin{document}

\title{Parametric self pulsing in a quantum  opto-mechanical  system.\footnote{Submitted to the special issue celebrating Paolo Tombesi's 70th birthday.}}

\author{C. A. Holmes, G. J. Milburn}
\affiliation{School of Mathematical and Physical Sciences, The University of Queensland, St Lucia, QLD 4072, Australia}

\begin{abstract}
We  describe an opto-mechanical system in which the coupling between optical and mechanical degrees of freedom takes the form of a fully quantised third-order parametric interaction. Two physical realisations are proposed: a harmonically trapped atom in a standing wave and the `membrane in the middle' model.  The dominant resonant interaction corresponds to a stimulated Raman process in which two phonons are converted into a single cavity photon. We show that this system can exhibit a stable limit cycle in which energy is periodically exchanged between optical and mechanical degrees of freedom. This is equivalently described as a parametric self-pulsing. 
\end{abstract}
\maketitle                   

\section{Introduction}
The ability to couple optical and mechanical degrees of freedom has driven innovation in many areas of modern physics, from the earliest demonstrations of laser cooling of atomic motion\cite{laser-cool} to the pioneering theory of Tombesi and co workers\cite{Tombesi} to recent experiments using macroscopic mirrors coupled cavity fields\cite{opto-mech}. In the latter example, most investigations are based on the radiation pressure coupling between the photon number in a cavity and the displacement of a mechanical resonator, typically a mirror,  which is linear in that displacement\cite{Wilson-Rae}. 

In this paper, by contrast, we describe a more complex dynamical situation in which the photon number in the cavity couples to the squared displacement of a mechanical resonator. We will consider two physical implementations, one based on harmonically trapped atoms interacting with a standing wave in a cavity via the off-resonant dipole interaction\cite{Domokos,Zippilli} , and the 'membrane in the middle' approach\cite{Thompson} .  This means that the cavity field experiences a phase modulation at twice the mechanical vibrational frequency. If this modulation frequency is larger than the cavity line-width, we can resolve these sidebands spectroscopically and drive the cavity on the first red sideband. In that case we can linearise the interaction with respect to the cavity field (not the mechanical system) and the dominant resonant terms then take the form of a fully quantised third-order parametric interaction. We show that this system can exhibit a stable limit cycle in which energy is periodically exchanged between optical and mechanical degrees of freedom. A semiclassical analysis, based on the centre manifold, enables us to calculate the amplitude and period of these oscillations. A first approach to a full quantum treatment enables us to calculate the phase diffusion constant on the limit cycle. 

Domokos and Ritsch\cite{Domokos} describe a system in which one or more atoms are trapped by the optical dipole force of a standing wave in an optical cavity and, as they move, modulate the standing wave amplitude.  Zippilli and Morigi\cite{Zippilli} consider a variation on this scheme in which the atoms are independently trapped by a harmonic potential provided by another focussed optical dipole trap or an ion trap.  We consider a similar model but, like Domokos and Ritsch, include the back-action of atoms on the standing wave of the optical cavity. If the atoms are trapped at a node of the cavity field, the Hamiltonian for the system is 
\begin{equation}
H=\hbar \omega_c a^\dagger a +\hbar\nu b^\dagger b+\hbar(\epsilon_c^* a e^{i\omega t}+\epsilon_c a^\dagger e^{-i\omega t})+\hbar G a^\dagger a (b+b^\dagger)^2 
\label{ham}
\end{equation}
where $\omega_c$ is the cavity optical frequency, $\nu$ is the mechanical trapping frequency,  $\epsilon_c$ represents a coherent cavity driving field at frequency $\omega$, and the coupling constant is given by
\begin{equation}
G=\frac{\eta^2 g^2}{\Delta}
\end{equation}
where $g$ is the single photon Rabi frequency for the atomic dipole transition, $\Delta$ is the detuning between the atomic dipole resonance and the cavity field. The Lamb-Dicke parameter is 
\begin{equation}
\eta=k\sqrt{\frac{\hbar}{2m\nu}}
\end{equation}
and $k$ is the wave number for the cavity field.    Typically the atom is harmonically bound with a  typical trapping frequency in the range $\nu \sim 10^4 - 10^5$ Hz. This is much smaller than the optical frequency, but can be of the order of the detuning between the cavity field and the cavity driving field. 
In an interaction picture at frequency $\omega$ the Hamiltonian may be written as 
\begin{equation}
H=\hbar \delta a^\dagger a +\hbar\nu b^\dagger b+\hbar(\epsilon_c^* a +\epsilon_c a^\dagger )+\hbar G a^\dagger a (b+b^\dagger)^2
\end{equation}
where $\delta=\omega_c-\omega$.

In the absence of the opto-mechanical coupling ($G=0$), the cavity field will reach a steady state which is a coherent state with amplitude $\langle a\rangle_{ss}=\bar{\alpha}$ with 
\begin{equation}
\bar{\alpha}=\frac{-i\epsilon_c}{\kappa/2-i\delta}
\end{equation}
where $\kappa$ is the photon decay rate for the cavity field. In the presence of weak opto-mechanical coupling we proceed by transforming to the {\em displacement picture} to remove the coherent state component of the optical mode,
\begin{equation}
\bar{H}=D^\dagger(\bar{\alpha})HD(\bar{\alpha})
\end{equation}
Then
\begin{equation}
\bar{H}=\hbar \delta a^\dagger a +\hbar\nu b^\dagger b+\hbar G|\bar{\alpha}|^2(b+b^\dagger)^2+\hbar G (\bar{\alpha}a^\dagger +\bar{\alpha}^* a)(b+b^\dagger)^2
\end{equation} 
We will further assume that the mechanical frequency is larger than the cavity line-width, $2\nu>\kappa$, so that the optical cavity can be driven on resolved sidebands. We then  transform to an interaction picture via the unitary transformation $U=\exp[-i\delta t a^\dagger a-i\nu tb^\dagger b]$, and assume the parametric resonance condition
\begin{equation}
\delta=\pm2\nu
\end{equation}
Keeping only resonant terms we have two different kinds of resonant interactions
\begin{equation}
H_r=\hbar G (\bar{\alpha}a^\dagger b^2+\bar{\alpha}^* ab^{\dagger 2})+2\hbar G |\bar{\alpha}|^2b^\dagger b
\label{red}
\end{equation}
when the laser driving the cavity is tuned to the red (lower frequency) of the cavity frequency $\omega=\omega_c-2\nu$, and
\begin{equation}
H_b=\hbar G (\bar{\alpha}a^\dagger b^{\dagger 2}+\bar{\alpha}^* a b^2)+2\hbar G |\bar{\alpha}|^2b^\dagger b
\end{equation}
when the laser driving the cavity  is tuned to the blue (higher frequency) of the cavity frequency $\omega=\omega_c+2\nu$. These Hamiltonians describe two-phonon Raman processes. In what follows we will absorb the Stark shift of the mechanical frequency (proportional to $|\bar{\alpha}|^2$) into the definition of the mechanical frequency.

Thompson et al.\cite{Thompson} describe an opto-mechanical system comprising an optical cavity containing a thin dielectric membrane of SiN placed between rigid high-finesse mirrors. The Hamiltonian describing the interaction between the optical mode and the mechanical displacement of the membrane is given by \cite{Thompson}
\begin{equation}
H=\hbar\omega_{c}(x)a^\dagger a+\hbar\nu b^\dagger b+\hbar(\epsilon_c^* a e^{i\omega t}+\epsilon_c a^\dagger e^{-i\omega t})
\end{equation}
where $a,a^\dagger$ refer to the optical mode and $b,b^\dagger$ refer to the mechanical resonator and we again include a coherent driving field on the cavity and
\begin{equation}
x=\sqrt{\frac{\hbar}{2\nu m}}(b+b^\dagger)
\end{equation}
If the membrane is placed at $x=0$, an extremum of $\omega_c$, we approximate
\begin{equation}
H=\hbar\omega_{c}a^\dagger a+\hbar\nu b^\dagger b+\hbar(\epsilon_c^* a e^{i\omega t}+\epsilon_c a^\dagger e^{-i\omega t})
+\hbar G a^\dagger a(b+b^\dagger)^2
\label{ham2}
\end{equation}
where $\omega_c=\omega_c(0)$ and
\begin{equation}
G=\frac{\hbar}{4\nu m}\left . \frac{\partial^2\omega_c}{\partial x^2}\right |_{x=0}
\end{equation}
The Hamiltonian in Eq.(\ref{ham2}) is the same form as the previous model in Eq. (\ref{ham}). In a similar way we can now consider two resonance conditions depending on whether the cavity field is driven on the red side band or the blue side band. In this paper we will consider only the red sideband case given by Eq. (\ref{red}). 

The possibility of realising two phonon Raman processes in this system has been noted by Thompson et al\cite{Thompson}. For the parameters given in Thompson et al.\cite{Thompson}, $\kappa\sim 10^4-10^5\ {\rm s}^{-1},\ \ \gamma\sim 10^2{\rm s}^{-1}$ while the mechanical frequency is $\nu\sim 8\times 10^{5}{\rm s}^{-1}$.  Thus it should be possible to achieve the resolved side band regime for an optical cavity of sufficiently high finesse.

\section{Response to mechanical driving}
We consider a cavity mode coupled to a mechanical resonator via the interaction Hamiltonian in Eq. (\ref{red}). This describes a situation in which two times  the mechanical frequency, $2\nu$, is larger than the cavity decay rate, $\kappa$, so that we can drive the cavity on the first red sideband. We now consider the response of the cavity field to a coherent driving force on the mechanical resonator. This is the analogue of the nonlinear optical process of second harmonic generation in which excitations in the driven mode at frequency $\nu$ are converted to a single excitation in the un driven mode at frequency $2\nu$. In the case considered here however the transitions are Raman-like processes in which two phonons are up-converted into one cavity photon by absorbing cavity pump photons at the sideband frequency $\omega_c-2\nu$.   

Moving to an interaction picture and neglecting counter rotating terms in the interaction between cavity field-mechanical resonator, we obtain the interaction picture Hamiltonian,
\begin{equation}
H_I(t)=\hbar\chi\left (a^\dagger  b^2 + a(b^\dagger)^2\right )-\hbar\epsilon(b+b^\dagger)  \label{eq:1}
\end{equation}
where, for simplicity, we assume that we can choose the phase of the cavity sideband driving, $\epsilon_c$,  to make $\bar{\alpha}$ real and we define
\begin{equation}
\chi=G\bar{\alpha}
\end{equation}
Note that while $G$ may be small $G\bar{\alpha}$ can be large. This is very similar to the strong coupling limit recently achieved for a linear opto-mechanical system\cite{Aspelmeyer}

In addition to the explicit optical and mechanical degrees of freedom included in the Hamiltonian, we need to model dissipation and associated noise sources. The resulting master equation ( in the interaction picture) for the total state of optical plus mechanical oscillators is then\cite{Walls_Milb}
 \begin{equation}
\frac{d\rho}{dt}  =  -\frac{i}{\hbar}[H_I(t),\rho]+\kappa{\cal D}[a]\rho+\gamma(\bar{n}+1){\cal D}[b]\rho+\gamma\bar{n}{\cal D}[b^\dagger]
  \label{ME}
\end{equation}
where $\kappa,\ \ \gamma$ are the amplitude damping rates of the optical and mechanical oscillators respectively, $\bar{n}$ is the mean bosonic occupation numbers for the mechanical bath.  The super operator ${\cal D}$ is defined by
\begin{equation}
{\cal D}[A]\rho=A\rho A^\dagger -\frac{1}{2}\left (A^\dagger A \rho +\rho A^\dagger A\right )
\end{equation}

The master equation, Eq.(\ref{ME}), has been previously used to describe sub-second harmonic generation for two quantised fields interacting in a medium with a significant second order optical nonlinearity\cite{Walls_Milb}. The first step to understanding the behavior is to consider the semiclassical equations of motion. These exhibit a rich behavior including fixed points, and limit cycles. We will follow the treatment of Drummond et al. \cite{DMcNW}. We proceed by first analyzing the classical equations of motion for the system.

\subsection{Semiclassical dynamics}

The classical non-linear equations of motion for the red sideband interaction are,
\begin{equation}
\left\{ \begin{array}{l}
\dot{\alpha} = \imath \chi \beta^2 - \frac{\kappa}{2} \alpha,\\
\dot{\beta} = 2 \imath \chi \beta^* \alpha - \imath \epsilon - \frac{\gamma}{2} \beta,
\end{array} \right. \label{e01}
\end{equation}
together with their complex conjugate counterparts. Here $\alpha$ and $\beta$ are the complex amplitudes of the cavity field and the mechanical oscillator respectively.  Setting $\chi=1$, as it simply scales time, and rewriting in real form gives
\begin{eqnarray}
 \dot{\beta_{r}} & = &  2(\beta_{i} \alpha_{r}-\beta_{r}\alpha_{i}) -\frac{\gamma}{2}\beta_{r},\\
  \dot{\beta_{i}} &  = &  2(\beta_{r}  \alpha_{r} + \beta_{i} \alpha_{i}) - \frac{\gamma}{2}\beta_{i} -\epsilon\\
  \dot{\alpha_{r}} & = &  -2\beta_{r}\beta_{i}  -  \frac{\kappa}{2}\alpha_{r},\\
   \dot{\alpha_{i}} & = &  \beta_{r}^2-\beta_{i}^2-\frac{\kappa}{2}\alpha_{i}
   \label{real-imag}
   \end{eqnarray}
This system has  one critical point existing for all $\epsilon$.

The real parts of the variables at the
 critical point 
  are zero; $\beta_{r0}=0,\,\alpha_{r0}=0$. The imaginary part of $\beta$ is then given by the solutions to the following cubic 
\begin{equation}
 \frac{4}{\kappa} \beta_{i0}^3 +  
   \frac{\gamma}{2}\beta_{i0} +\epsilon =0
   \label{critical-eqn}
   \end{equation}
  Note that if $\epsilon$ is assumed to be positive then $\beta_{i0}$ is negative. (The other case ($\epsilon<0,\,\beta_{i0}>0$) can be obtained from symmetry.)
 The imaginary part of   $\alpha$ is given in terms of the solution 
on the cubic; $\displaystyle{\alpha_{i0} =  - \frac{2\beta_{i0}^2}{\kappa}.}$

 The linearized matrix  about this critical point is
   $$\left(\begin{array}{ccccrrrr} -2\alpha_{i0}-\frac{\gamma}{2}& 
   0&2\beta_{i0}&0\\ 0& 2\alpha_{i0}-\frac{\gamma}{2}&0& 2\beta_{i0}\\
   -2\beta_{i0}&0&-\frac{\kappa}{2}&0\\
   0&-2\beta_{i0}&0&-\frac{\kappa}{2}\end{array}\right)$$
   Now at a possible Hopf bifurcation two of 
   the eigenvalues of this linearized matrix are pure imaginary.
   This  can occur here  when
   $\alpha_{i0} = -\frac{\kappa+\gamma}{4}$, which gives
\begin{equation}
|\beta|^2=\frac{\kappa(\kappa+\gamma)}{8},\hspace{1cm} 
\epsilon = \epsilon_{h} = 
  -\frac{\beta_{i0}(\kappa+2\gamma)}{2} = \frac{\sqrt{\kappa(\kappa+\gamma)}(\kappa+2\gamma)}{4\sqrt{2}}
\end{equation}
    Note that for $\kappa \gg \gamma$ then $\epsilon_{h} \approx \frac{\kappa^2}{4\sqrt{2}}$. 

We have assumed a situation where the Hopf bifurcation is analyzed as 
    $\epsilon$ is varied away from $\epsilon_h$ and the other parameters are fixed. So the calculation of the center manifold at the bifurcation is performed for  $\epsilon= \epsilon_h$.
 The linearized matrix  at the Hopf bifurcation is
   $$\left(\begin{array}{ccccrrrr}\frac{\kappa}{2}&
   0&2\beta_{i0}&0\\ 0&-\gamma-\frac{\kappa}{2}&0& \beta_{i0}\\
   -2\beta_{i0}&0&-\frac{\kappa}{2}&0\\
      0&-2\beta_{i0}&0&-\frac{\kappa}{2}\end{array}\right)$$
where $\beta_{i0} =\beta_{i0h}=- \sqrt{\frac{\kappa(\kappa+\gamma)}{8}}. $ This has eigenvalues $$
\lambda_{1\pm} = \pm i \frac{\sqrt{\kappa(\kappa+2\gamma)}}{2}, \hspace{1cm}
\lambda_{2\pm} = -\frac{\kappa+\gamma}{2} \pm i \frac{\sqrt{2\kappa(\kappa+\gamma)-\gamma^2}}{2}$$ 

The equations of motion after perturbing off the critical point at\\
 $(0,\,\beta_{i0h},\,0,\,\alpha_{2i0}=\frac{2\beta_{i0h}^2}{\kappa})$,
    are
\begin{equation}
\left( \begin{array}{l} \dot{\beta_{r}}\\\dot{\alpha_{r}}\\
 \dot{\beta_{i}}\\ \dot{\alpha_{i}}\end{array}\right)
 = \left( \begin{array}{ccccrrrr}\frac{\kappa}{2}&2\beta_{i0h}&
    0&0\\ -2\beta_{i0h}&-\frac{\kappa}{2}&0&0\\
    0&0&-\gamma-\frac{\kappa}{2}& 2\beta_{i0h}\\
             0&0&-2\beta_{i0h}&-\frac{\kappa}{2}\end{array}\right)
\left(\begin{array}{c}\beta_{r}\\\alpha_{r}\\\beta_{i}\\\alpha_{i}\end{array}\right) + 
 \left( \begin{array}{c}
 2(\beta_{i} \alpha_{r}-\beta_{r}\alpha_{i}) \\
 -2\beta_{r}\beta_{i}\\
 2(\beta_{r}  \alpha_{r} + \beta_{i} \alpha_{i})\\
	      \beta_{r}^2-\beta_{i}^2\end{array}\right)
	      \end{equation}

The center eigenspace is the $(\beta_{r},\,\alpha_{r})$ space and the center manifold is tangent to the $(\beta_{r},\,\alpha_{r})$ space.
Say it is given, at least locally by $\beta_{i} = h_1(\beta_{r},\,\alpha_{r})$ and $\alpha_{i} = h_2(\beta_{r},\,\alpha_{r})$, then to lowest order in $\beta_{r}$ and $\alpha_{r}$  the center manifold is given by
 \begin{eqnarray}
\beta_{i} & = &  h_1(\beta_{r},\,\alpha_{r})\approx A_1 \beta_{r}^2 + B_1 \beta_{r}\alpha_{r} + C_1\alpha_{r}^2\\
\alpha_{i} & = & h_2(\beta_{r},\,\alpha_{r}) \approx A_2 \beta_{r}^2 + B_2\beta_{r}\alpha_{r} + C_2\alpha_{r}^2
\end{eqnarray}
Substituting back into the equations of motion gives
values for $A_i,\,B_i,\,C_i$. 
$$ A_1 = \frac{2-2\sqrt{2}\sqrt{\kappa(\kappa+\gamma)}\kappa(27\kappa^3+92\gamma\kappa^2+96\kappa\gamma^2+16\gamma^3)}{D}$$$$
B_1=\frac{4\kappa^2(11\kappa^2+34\kappa\gamma+32\gamma^2)(2\gamma+3\kappa)}{D} $$$$C_1 = \frac{-(4(2\gamma+3\kappa))\sqrt{2}\sqrt{\kappa(\kappa+\gamma)}\kappa(\kappa^2+2\kappa\gamma+4\gamma^2)}{D}
$$ $$
A_2 = \frac{2\kappa(5\kappa^3+24\gamma\kappa^2+32\kappa\gamma^2+16\gamma^3)(2\gamma+3\kappa)}{D}$$$$
B_2=\frac{-(8(2\gamma+3\kappa))\sqrt{2}\sqrt{\kappa(\kappa+\gamma)}\kappa^2(2\kappa+5\gamma)}{D} $$$$C_2 = \frac{(8(\kappa+2\gamma))(5\kappa+2\gamma)\kappa(\kappa+\gamma)(2\gamma+3\kappa)}{D}
$$
where $D=(\kappa^2(4\gamma+3\kappa)(32\gamma^3+96\kappa\gamma^2+72\kappa^2\gamma+17\kappa^3))$.
Thus the equations of motion on the center manifold,
for $\epsilon =\epsilon_{h}$, are
\begin{eqnarray}
\left( \begin{array}{l} \dot{\beta_{r}}\\\dot{\alpha_{r}}
  \end{array}\right)
   & = & \left( \begin{array}{ccrr}\frac{\kappa}{2}&2\beta_{i0h}
       \\ -2\beta_{i0h}&-\frac{\kappa}{2}
	              \end{array}\right)
	\left(\begin{array}{c}\beta_{r}\\\alpha_{2r}\end{array}\right)\nonumber\\
	& & +  \left( \begin{array}{c}
  2(-A_2\beta_{r}^3 + (A_1-B_2) \alpha_{r}\beta_{r}^2+(B_1-C_2)\beta_{r}\alpha_{r}^2 + C_1 \alpha_{r}^3) \\
			   -2\beta_{r}(A_1 \beta_{r}^2 + B_1 \beta_{r}\alpha_{r} + C_1\alpha_{r}^2)
			                  \end{array}\right)
			                  \end{eqnarray}
where $\beta_{i0h} =-\sqrt{\frac{\kappa(\kappa+\gamma)}{8}}. $

Transforming to normal form
\begin{equation}
\left (\begin{array}{c} \beta_{r}\\\alpha_{r}\end{array}\right) =
\left(\begin{array}{ccrr} 0& 2 \beta_{i0h}\\\frac{\sqrt{\kappa(\kappa+2\gamma)}}{2}&-\frac{\kappa}{2}\end{array}\right)
\left(\begin{array}{c}u\\v\end{array}\right)
\label{normal-form}
\end{equation}
which has the inverse 
\begin{eqnarray}
u & = & \frac{2\alpha_{r}}{\sqrt{\kappa(\kappa+2\gamma)}} + \frac{\kappa \beta_{r}}{2 \sqrt{\kappa(\kappa+2\gamma)} \beta_{i0h}},\\
v & = &  \frac{ \beta_{r}}{2\beta_{i0h}}.
\label{normal-form-inv}
\end{eqnarray}
The  equations of motion  in the normal form variables are
$$\dot{u} =  -\frac{\sqrt{\kappa(\kappa+2\gamma)}}{2} v  + \bar{A_1}u^3 
+\bar{B_1}u^2 v  +\bar{C_1}u v^2 
+\bar{D_1}v^3$$
$$\dot{v} = \frac{\sqrt{\kappa(\kappa+2\gamma)}}{2}u  + \bar{A_2}u^3 
+\bar{B_2}u^2 v  +\bar{C_2}u v^2 
+\bar{D_2}v^3$$
 where $ \bar{A_i},\,\bar{B_i},\,\bar{C_i},\,\bar{D_i}$ are functions of $\kappa$ and $\gamma$ and can be calculated from the equations above. 
 
 These equations are valid for $\epsilon =\epsilon_{h}$, however they can be extended to include the case where $\epsilon =\epsilon_{h}+\Delta \epsilon$, by considering the variation in the  trace of the linearized matrix with
 $\epsilon$.
 Recall that the linearized matrix for $(\beta_{r},\,\alpha_{r})$ at the critical point is
$$\left(\begin{array}{ccrr} -2\alpha_{i0}-\frac{\gamma}{2}& 
   2\beta_{i0}\\ 
   -2\beta_{i0}&-\frac{\kappa}{2}
   \end{array}\right)$$
   where $\displaystyle{\alpha_{i0} =  - \frac{2\beta_{i0}^2}{\kappa}.}$
The trace of this matrix is a function of $\epsilon$; 
 $$ trace(\epsilon) = - 
\frac{2(\beta_{i0}(\epsilon)^2)}{\kappa} -\frac{\kappa+\gamma}{2}
$$ since $\beta_{i0}$ must also be treated as a
 function of $\epsilon$ via the cubic
$$ \hspace{3mm} \frac{4}{\kappa} \beta_{i0}^3 
   + \frac{\gamma}{2}\beta_{i0} +\epsilon =0$$
   To a first approximation in $\Delta \epsilon$
   $$ trace(\epsilon) =  trace(\epsilon_{h} ) + \frac{\partial trace}{\partial \epsilon}|_{(\epsilon= \epsilon_{h})} \Delta \epsilon$$
   On calculation
   $$\frac{\partial trace}{\partial \epsilon}|_{(\epsilon= \epsilon_{h})}
   = \frac{\sqrt{8\kappa(\kappa+\gamma)}}{3\kappa+4\gamma} $$
   which  means that the linear system for $\epsilon \approx \epsilon_{h}$
   in polar coordinates is
   $$ \dot{r} = d \Delta \epsilon r \hspace{10mm} \mbox{where} \hspace{3mm}
   d = \frac{\sqrt{8\kappa(\kappa+\gamma)}}{\kappa(3\kappa+4\gamma)}$$

Finally the effect of the nonlinear terms  is to include an $r^3$ term, whose coefficient is  a function of the nonlinear coefficients given above(for details see \cite{Glendinning})
$$a = -\frac{\kappa^2(\kappa+\gamma)(99\kappa^4+490\gamma\kappa^3+808\kappa^2\gamma^2+512\kappa\gamma^3+128\gamma^4)}{4(128\kappa^2\gamma^4+480\kappa^3\gamma^3+51\kappa^6+284\kappa^5\gamma+576\kappa^4\gamma^2)}$$
Note that for $\kappa \gg \gamma$ then $ a\approx -\frac{33\kappa}{68}.$

Since $d>0$ and $a<0$ there is a supercritical Hopf bifurcation creating a stable limit cycle, which exists for $\Delta$  small and positive.
In polar coordinates this gives the approximate system for the radial equation as
$$\dot{r}= d\Delta\epsilon r + a r^3 \hspace{10mm} 
\mbox{which has non zero solution} \hspace{5mm} r^2 = \frac{d\Delta\epsilon}{a}.$$
This gives the amplitude of the limit cycle: $\displaystyle{A = \sqrt{\frac{d\Delta\epsilon}{a}}}.$
 The frequency of the resulting periodic orbit is
 \begin{equation}
  \omega_h=\frac{\sqrt{\kappa(\kappa+2\gamma)}}{2}
  \label{hopf-freq}
  \end{equation}
For $\kappa \gg \gamma$ the approximate amplitude and frequency of the resulting oscillation are 
\begin{equation}
A \approx \frac{1}{\kappa}\sqrt{\frac{136\sqrt{2}\Delta \epsilon }{99}}\hspace{2cm}
\omega_h\approx \frac{\kappa}{2}.
\label{lc-amp}
\end{equation}
However caution should be used when considering this equation as it is the amplitude in the transformed variables. As we will see later on the amplitude in the original variables is multiplied by $\kappa$. The solution on the limit cycle  in normal form variables is 
\begin{eqnarray}
 u & = &  A \sin ( \omega_h t + \phi)\\
 v & = &  A \cos( \omega_h t + \phi).
 \end{eqnarray}
In the original variables this becomes
\begin{eqnarray} \beta_{r} &=& 2\beta_{i0h} A \cos( \omega_h t + \phi)\\
 \alpha_{r} &=& \omega_h A \sin ( \omega_h t + \phi) -
 \frac{\kappa A}{2}\cos(\omega_h t + \phi)
\end{eqnarray}
and
$$\beta_{i} = \beta_{i0}+h_1(2\beta_{i0h} A \cos( \omega_h t + \phi),\,\omega_h A \sin ( \omega_h t + \phi) -
 \frac{\kappa A}{2}\cos(\omega_h t + \phi))$$ 
 $$\alpha_{i} = \alpha_{i0}+h_2(2\beta_{i0h} A \cos( \omega_h t + \phi),\,\omega_h A \sin ( \omega_h t + \phi) -
 \frac{\kappa A}{2}\cos(\omega_h t + \phi))$$ 

However it is important to note that the functions  $ h_j $ are nonlinear.
So that  near the bifurcation, that is for $\Delta{\epsilon}$ small,
$$ \alpha_{r} ,\,\beta_{r}= O(\sqrt{\Delta{\epsilon}}) \hspace{10mm}\mbox{but}
\hspace{4mm} \alpha_{i} = \alpha_{i0} + O(\Delta{\epsilon})\hspace{4mm}\mbox{and}\hspace{4mm}\beta_{i}=\beta_{i0}+ O(\Delta{\epsilon})$$
In other words it is consistent to consider an approximation only to order
$\sqrt{\Delta{\epsilon}}$ in which case
\begin{eqnarray}
\beta_{r} &=& 2\beta_{i0h} A \cos( \omega_h t + \phi)\\
 \alpha_{r} &=& \omega A \sin ( \omega_h t + \phi) -
 \frac{\kappa A}{2}\cos(\omega_h t + \phi)\\
\beta_{i}  &=& -\sqrt{\frac{\kappa(\kappa+\gamma)}{8}}-\frac{2\Delta \epsilon}{3\kappa+4\gamma}\\
\alpha_{i}  &=& -\frac{\kappa+\gamma}{4}-\frac{2\sqrt{2\kappa(\kappa+\gamma)}\Delta \epsilon}{\kappa(3\kappa+4\gamma)} 
\end{eqnarray}

\begin{figure}[h!]
\includegraphics[scale=0.7]{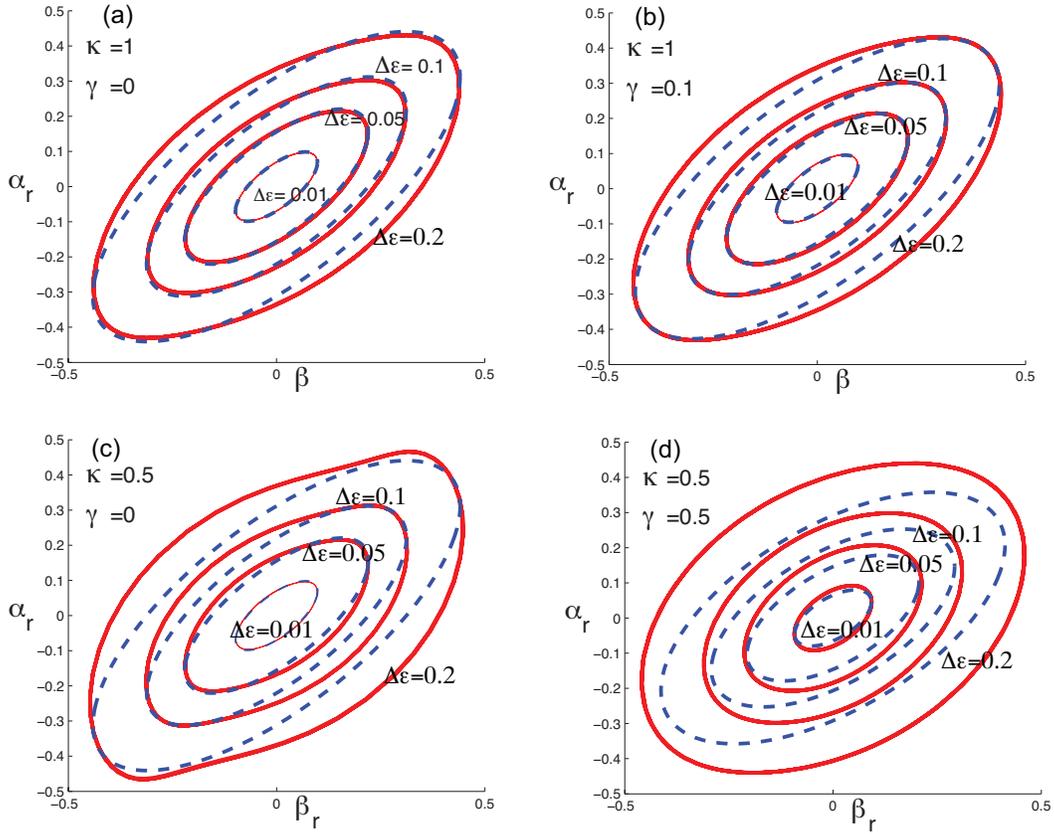}
\caption{A comparison of the approximate limit cycles (dashed) with the numerical solutions (solid) for various values of the parameters, (a)$\kappa=1.0,\gamma=0.0$ (b)$\kappa=1.0,\gamma=0.1$ (c) $\kappa=0.5, \gamma=0.0$ (d) $\kappa=0.5,\gamma=0.5$} \label{limit-cycles} \end{figure}

In figure [\ref{limit-cycles}] the actual limit cycles are shown in red and the approximations from center manifold theory in blue dashed lines. The limit cycles are shown for various values of the bifurcation parameter $\Delta \epsilon$ and specific values of $\kappa$ and $\gamma$. 

\vskip 1 truecm

\section{\label{sec:three} Quantum Noise. }
The semiclassical analysis gives us an understanding of the relevant dynamical structures including fixed points and limit cycles. However to understand the role of noise in the system it is necessary to proceed to a full quantum treatment. While an exact time dependent solution is not available, considerable insight can be gained by considering the linearised theory near the fixed points. Such an approach must necessarily fail at the fixed points themselves as by definition the linear terms in the dynamical equations vanish at these points, and the eigenvalues of the linearised analysis go to zero. Nonetheless such a  approach has proved very effective in understanding similar systems in quantum optics and we except the same to be true for opto-mechanical systems.   

We follow the treatment of Drummond et al. \cite{DMcNW} based on a quantum stochastic differential equations in the generalized P-representation. In this approach the quantum operators $a,a^\dagger,b, b^\dagger$ are replaced by independent complex variables, $\alpha,\alpha^\dagger,\beta,\beta^\dagger$. Note that the dimension of the phase space has doubled in this representation. This is the price one must pay to follow quantum noise for anti-normally ordered moments. The semiclassical equations arise by neglecting noise and restricting to the subspace $\alpha^\dagger\rightarrow \alpha^*,\beta^\dagger\rightarrow \beta^*$. We will only consider the case of zero temperature in this paper, $\bar{n}=0$. 

\begin{eqnarray*}
\frac{\partial}{\partial t} \left( \begin{array}{c} \beta \\ \beta^\dagger \end{array} \right) &=& \left(\begin{array}{l} 2 \imath \chi \beta^\dagger \alpha - \imath \epsilon - \frac{\gamma}{2}\beta \\ -2 \imath \chi \beta \alpha^\dagger + \imath \epsilon - \frac{\gamma}{2} \beta^\dagger  \end{array}   \right) + \left(\begin{array}{c c} 2 \imath \chi  \alpha & 0 \\ 0 & -2 \imath \chi \alpha^\dagger \end{array} \right)^{1/2} \left( \begin{array}{c} \eta_1 (t) \\ \eta_1^\dagger (t) \end{array} \right) \\
\frac{\partial}{\partial t} \left( \begin{array}{c} \alpha \\ \alpha^\dagger \end{array} \right) &=& \left(\begin{array}{l} \imath \chi \beta^2  - \frac{\kappa}{2}\alpha \\ - \imath \chi (\beta^\dagger)^2 - \frac{\kappa}{2} \alpha^\dagger  \end{array}   \right) 
\end{eqnarray*} 
If we linearise around the semiclassical fixed points (i.e. prior to the Hopf bifurcation $\epsilon<\epsilon_h$), the stochastic equations of motion are
\begin{eqnarray}
\frac{\partial}{\partial t} \left( \begin{array}{c} \delta \beta \\ \delta \beta^\dagger \\ \delta \alpha \\ \delta \alpha^\dagger  \end{array} \right) &=& \left( \begin{array}{cccc} -\frac{\gamma}{2} & 2 \imath \chi \alpha_0 & 2 \imath \chi \beta^*_{0} & 0 \\ - 2 \imath \chi \alpha^*_{0} & -\frac{\gamma}{2} & 0 & - 2 \imath \chi \beta_0 \\  2 \imath \chi \beta_0 & 0 & -\frac{\kappa}{2} & 0 \\ 0 & -2 \imath \chi \beta^*_{0} & 0 & -\frac{\kappa}{2}   \end{array} \right) \left( \begin{array}{c} \delta \beta \\ \delta \beta^\dagger \\ \delta \alpha \\ \delta \alpha^\dagger  \end{array} \right)  \nonumber \\
 & &\ \ \ \ \ \ + \left( \begin{array}{cccc} 2 \imath \chi  \alpha_0 & 0 & 0 & 0 \\ 0 & -2 \imath \chi \alpha^*_{0} & 0 & 0 \\ 0&0&0& 0  \\ 0&0& 0& 0 \end{array} \right)^{1/2} \left( \begin{array}{c} \eta_1 (t) \\ \eta_1^\dagger (t) \\ \eta_2(t) \\ \eta_2^\dagger(t) \end{array} \right)  ,\nonumber 
\end{eqnarray}
which we write as 
\begin{equation}
\frac{\partial}{\partial t} \left[ \delta \bm\alpha \right] = -A \left[ \delta \bm{\alpha} \right] + D^{1/2} \left[ \bm{\eta}(t) \right]  . \label{qu1}
\end{equation}
The steady state spectrum for normally ordered moments are defined by
\begin{equation}
S_{ij}(\omega)\frac{1}{2\pi}\int_{-\infty}^{\infty} e^{-i\omega\tau}\langle \alpha_i(t)\alpha_j(t+\tau)\rangle_{t\rightarrow\infty}\ d\tau
\end{equation}
which  is given in terms of the linearised drift ($A$) and diffusion ($D$) as\cite{Walls_Milb}
\begin{equation}
S(\omega) = \frac{1}{2 \pi} \left( \imath \omega I + A \right)^{-1} D \left(-\imath \omega I + A^{T}\right)^{-1}, 
\end{equation}
where $I$ is the identity matrix and the superscript $T$ denotes the transpose.

\begin{figure} [htp]
\begin{center}
\includegraphics[scale=1.0]{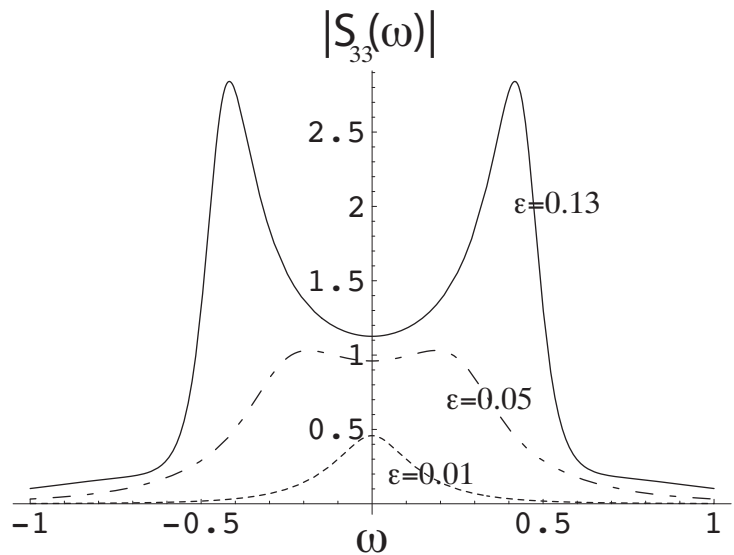}
\end{center}
\caption{\label{spectrum} Linearized spectrum for $|S_{33}(\omega)|$, for $\kappa=1.0,\gamma=0.1$, with  three different cases of $\epsilon= 0.01$, $\epsilon= 0.05$ and $\epsilon= 0.13$.} 
\end{figure}

Direct observations can be made on the output field from the cavity so we look for signatures of the Hopf bifurcation in the spectrum of the output field amplitude. In figure \ref{spectrum} we plot the noise spectrum for the optical field  amplitude for values of mechanical driving approaching the Hopf bifurcation. The spectrum becomes more sharply peaked on a frequency $\omega_h$ which is equal to the imaginary eigenvalue at the Hopf bifurcation, and given by Eq.(\ref{hopf-freq}). 
After the Hopf bifurcation, the real parts of $\alpha,\beta$ are no longer locked to zero but oscillate according to Eq.(36,37). The quantum noise will cause phase diffusion around the limit cycle. 

We can estimate this phase diffusion rate using the method of Zhu and Yu\cite{Zhu} for the response of a van der Pol oscillator to white noise in the limit $\kappa >>\gamma$. Using the linear transformation from $\alpha,\beta$ to $u,v$ variables to also transform the noise, we find that both $u$ and $v$ are subject to Gaussian white noise with diffusion constant approximately given by $s=1/\kappa$ close to the Hopf bifurcation. The phase on the limit cycle then undergoes a a diffusion process, 
\begin{equation}
\frac{d\phi}{dt}=\frac{1}{A}\sqrt{\frac{s}{2}}dW(t)
\end{equation}
where $dW$ is a Wiener process, and $A$ is the amplitude of the limit cycle given approximately by Eq.(\ref{lc-amp}).  The phase diffusion constant, for $\kappa>>\gamma$,  is thus
\begin{equation}
D_\phi=0.26\frac{\kappa}{\Delta\epsilon}
\end{equation}
Like a laser, the phase diffusion rate decreases as the limit cycle becomes bigger.

\vskip 1 truecm
\acknowledgments
This work has been supported by the Australian Research Council.

\end{document}